\documentclass[screen, nonacm]{acmart}
\AtBeginDocument{%
  \providecommand\BibTeX{{%
    \normalfont B\kern-0.5em{\scshape i\kern-0.25em b}\kern-0.8em\TeX}}}

\setcopyright{rightsretained}
\copyrightyear{2021}

\usepackage{graphicx}
\usepackage{array,booktabs,longtable,tabularx,subcaption,multirow}
\newcolumntype{Y}{>{\centering\arraybackslash}X}

\newcommand{\p}{\emph{p}$<$}
\newcommand{\pequal}{\emph{p}=}

\newcommand{\Z}{\emph{Z}=}
\newcommand{\F}[2]{\emph{F}$_{({#1},{#2})}$=}
\newcommand{\etasq}{$\eta_{p}^{2}$=}
\newcommand{\Friedman}{\emph{X}$^2$}

\newcommand{\base}{\emph{Physical}}
\newcommand{\none}{\emph{None}}

\newcommand{\cont}{\emph{Continuous}}
\newcommand{\inplace}{\emph{InPlace}}
\newcommand{\teleturn}{\emph{TeleTurn}}
\newcommand{\ttwo}{\emph{$22.5^{\circ}$}}
\newcommand{\ffive}{\emph{$45^{\circ}$}}
\begin{document}

\title{Augmenting Teleportation in Virtual Reality With Discrete Rotation Angles}

\author{Dennis Wolf}
\affiliation{
  \institution{Ulm University}
  \city{Ulm}
  \country{Germany}}
\email{dennis.wolf@uni-ulm.de}

\author{Michael Rietzler}
\affiliation{
  \institution{Ulm University}
  \city{Ulm}
  \country{Germany}}
\email{michael.rietzler@uni-ulm.de}

\author{Laura Bottner}
\affiliation{
  \institution{Ulm University}
  \city{Ulm}
  \country{Germany}}
\email{laura.bottner@uni-ulm.de}

\author{Enrico Rukzio}
\affiliation{
  \institution{Ulm University}
  \city{Ulm}
  \country{Germany}}
\email{enrico.rukzio@uni-ulm.de}

\renewcommand{\shortauthors}{Wolf et al.}

\begin{abstract}
Locomotion is one of the most essential interaction tasks in virtual reality (VR) with teleportation being widely accepted as the state-of-the-art locomotion technique at the time of this writing. A major draw-back of teleportation is the accompanying physical rotation that is necessary to adjust the users' orientation either before or after teleportation. This is a limiting factor for tethered head-mounted displays (HMDs) and static body postures and can induce additional simulator sickness for HMDs with three degrees-of-freedom (DOF) due to missing parallax cues. To avoid physical rotation, previous work proposed discrete rotation at fixed intervals (InPlace) as a controller-based technique with low simulator sickness, yet the impact of varying intervals on spatial disorientation, user presence and performance remains to be explored. An unevaluated technique found in commercial VR games is reorientation during the teleportation process (TeleTurn), which prevents physical rotation but potentially increases interaction time due to its continuous orientation selection. 
In an exploratory user study, where participants were free to apply both techniques, we evaluated the impact of rotation parameters of either technique on user performance and preference.
Our results indicate that discrete InPlace rotation introduced no significant spatial disorientation, while user presence scores were increased. 
Discrete TeleTurn and teleportation without rotation was ranked higher and achieved a higher presence score than continuous TeleTurn, which is the current state-of-the-art found in VR games. Based on observations, that participants avoided TeleTurn rotation when discrete InPlace rotation was available, we distilled guidelines for designing teleportation without physical rotation.
\end{abstract}



\begin{teaserfigure}
    \centering
    \begin{subfigure}[t]{.32\textwidth}
    \centering
        \includegraphics[height=.9\textwidth]{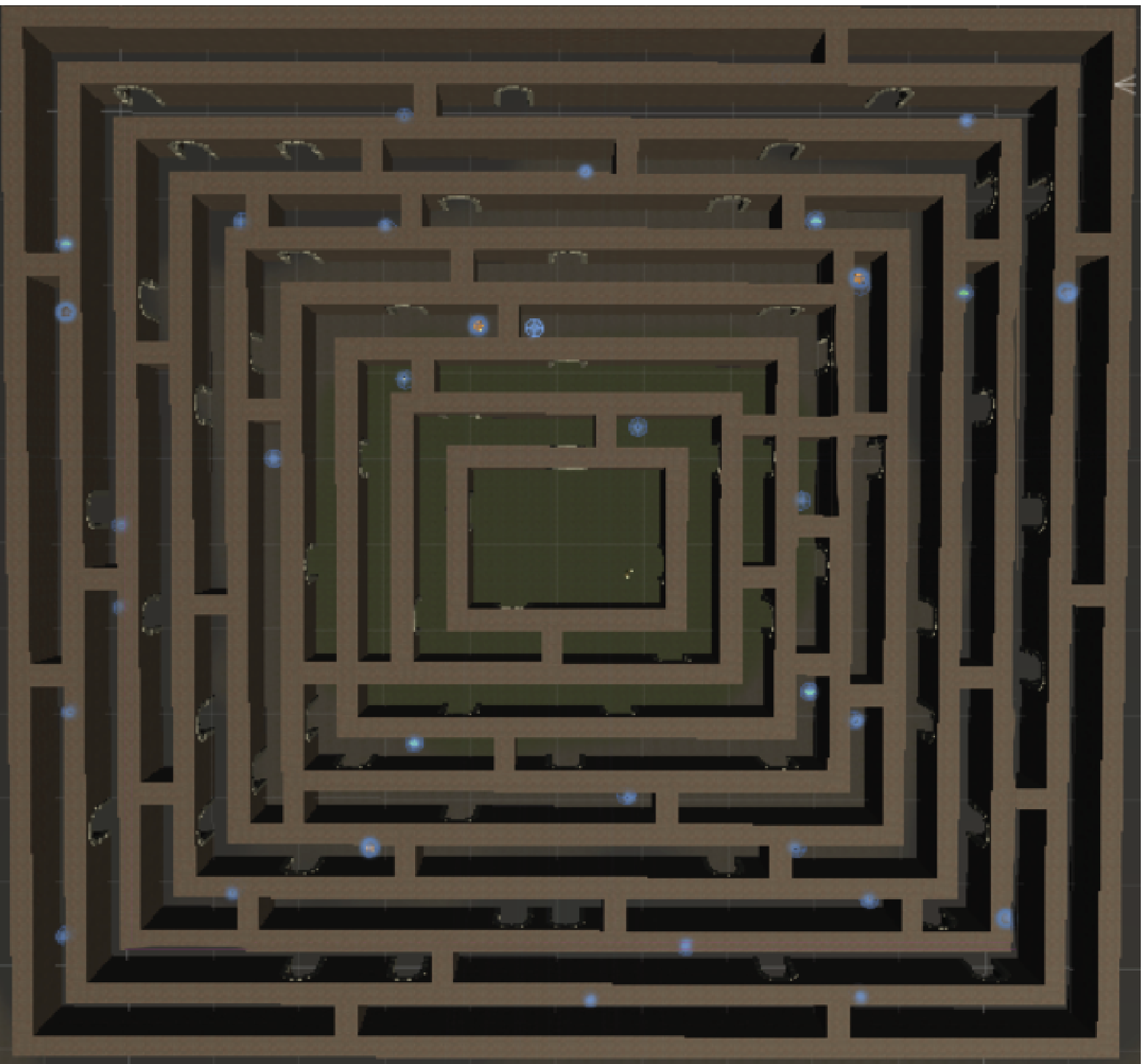}
        \subcaption{Study environment}
    \end{subfigure}
    \hfill
    \begin{subfigure}[t]{.32\textwidth}
    \centering
        \includegraphics[height=.9\textwidth]{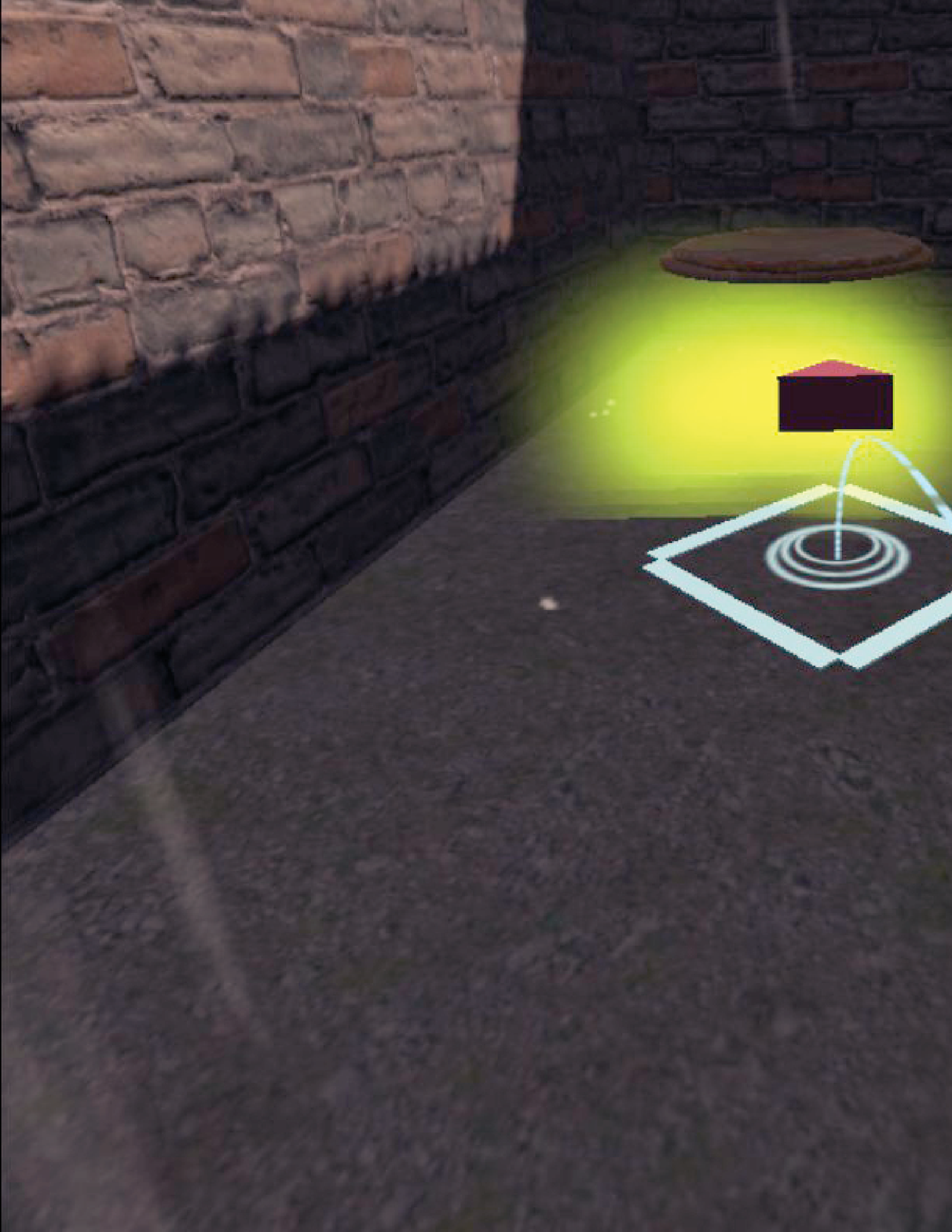}
        \subcaption{Collectible item}
    \end{subfigure}
    \hfill
    \begin{subfigure}[t]{.32\textwidth}
    \centering
        \includegraphics[height=.9\textwidth]{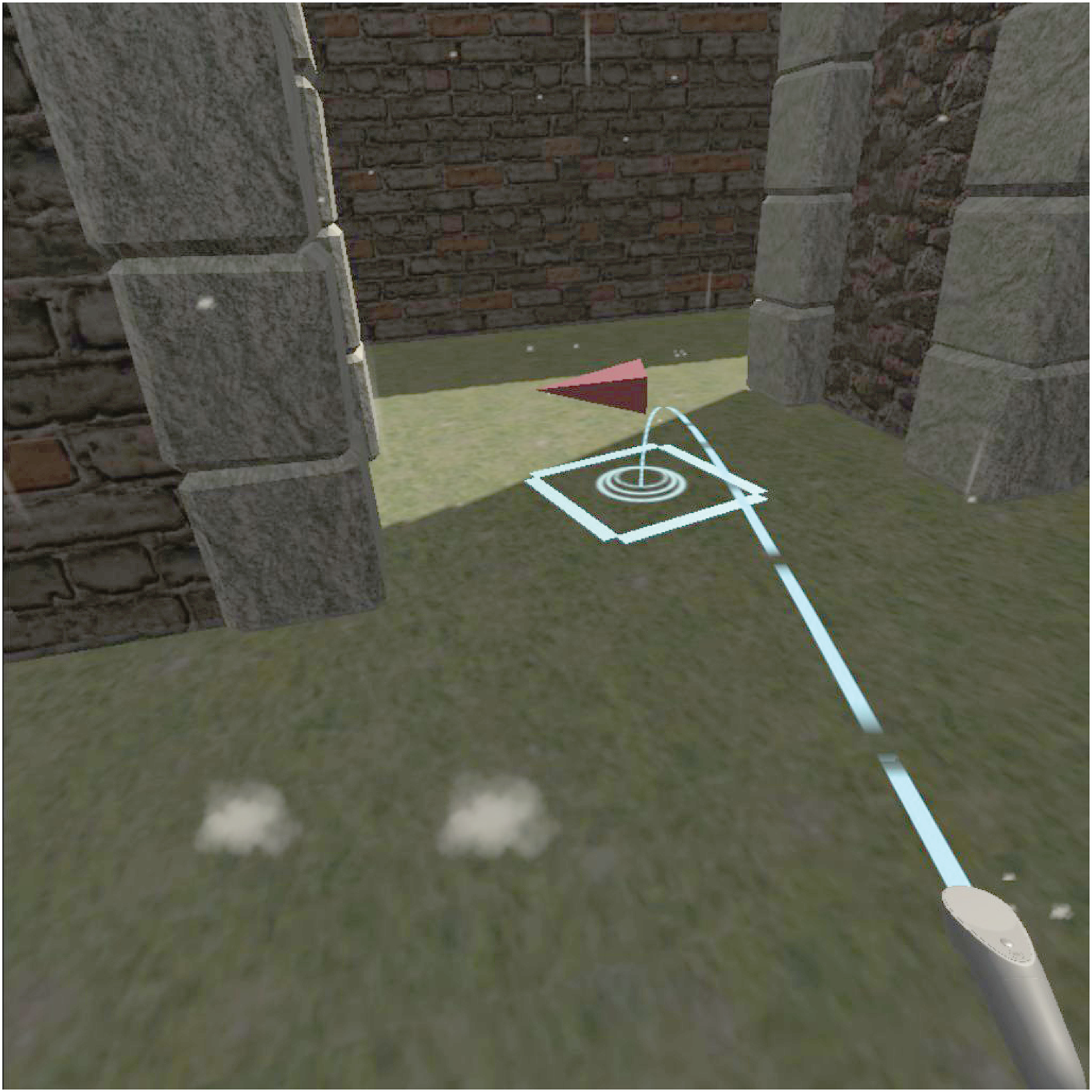}
        \subcaption{\teleturn~indicator}
    \end{subfigure}
    \caption{The study environment was designed as a labyrinth (a) with semi-randomly distributed items (b) to enforce rotations on the spot, i.e. \inplace~or during teleportations, i.e. \teleturn. In conditions with \teleturn~not set to \none, an arrow indicates the currently selected rotation direction (c).}
    \label{fig:discrete_labyrinth}
\end{teaserfigure}
\maketitle

\section{Introduction}
Virtual reality offers its users a potentially unlimited space to explore. Navigating this vast environment is a non trivial task and has been the focus of a large body of related work. Although walking in VR has been shown to be superior to walk-in-place and controller-based techniques~\cite{usoh1999walking}, spatial restrictions set limits to how far a user can walk in a straight line. To circumvent these restrictions, researchers propose to manipulate the virtual output presented to users redirecting them on a curved path (redirected walking with curvature gain)~\cite{razzaque2005redirected}. However, the large radius required to trick the users' perception renders a naive implementation of this technique unpractical for daily application~\cite{8613757}.
These limitations amongst others led to the rise of a seemingly ``unnatural'' and yet very efficient locomotion technique: teleportation (e.g., \cite{teleportation}). By simply pointing and pressing a button users can teleport to their destination in an instant. This technique allows to overcome spatial restrictions of a small tracking space and has been shown to induce less simulator sickness than controller-based continuous locomotion~\cite{Frommel:2017:ECL:3102071.3102082}.

A major drawback of state-of-the-art teleportation is a missing rotation component. Considering the tethered design of many commercial head-mounted displays (HMDs), users could get entangled by the cables during physical rotation. Furthermore, physical rotation might not be feasible (e.g., due to a physical impairment) or desirable depending on the context (e.g., during a sitting or lying body posture). In addition, the optical flow during physical rotation can cause a feeling of vection, i.e., the illusion of moving, and thus, induce simulator sickness~\cite{vection}. Especially low-class HMDs with only three degrees-of-freedom (DoF) are missing important parallax cues, which are necessary to compensate for postural changes during a physical rotation. The result is a mismatch between virtual and physical motion, which can induce simulator sickness and reduce the user experience~\cite{sensoryconflict}.

In an effort to extend teleportation-based locomotion by a rotation component, we identified two promising opportunities: rotation around the users' position (henceforth ``\inplace'') and rotation around the desired teleportation target (henceforth ``\teleturn''), which can be found in commercial VR games (e.g., Robo Recall~\footnote{\url{https://www.epicgames.com/roborecall/en-US/home}}).

To prevent any physical rotation, Sargunam et al. compared controller-based rotation techniques (\inplace) in terms of simulator sickness and disorientation and found discrete rotations at fixed intervals to induce the least simulator sickness~\cite{Sargunam:2018:EJC:3191801.3191815}. While they suggested an interval of $30^\circ{}$, this value stems from a preliminary study with no evaluation of the impact on user presence, disorientation, and preference. 

A commercially available, yet underexplored rotation technique is reorientation during the teleportation process (\teleturn). While selecting a teleportation destination, users can choose the new orientation that the user avatar should assume at the new destination point. This new orientation is typically selected via a continuous gesture on a joystick or touchpad and is displayed as an abstract arrow (see Figure~\ref{fig:discrete_labyrinth} c) or the users' avatar. However, a continuous gesture for orientation selection has been shown to slow down the interaction and reduce user acceptance of this technique~\cite{teleportation}. We are therefore interested in exploring the potential of a discrete version of \teleturn~to reduce interaction time and increase the overall usability. Furthermore, it is unclear how users would apply \inplace~and \teleturn~if both techniques are made available for direct comparison.

Results of our exploratory user study (N=12) indicate that users preferred discrete \inplace~rotation over discrete \teleturn~with the largest rotation parameter ($45^\circ{}$) being the favorite. However, discrete \teleturn~rotation showed an improvement over the current state-of-the-art, which is a continuous gesture for orientation selection.

Our contributions are therefore:
\begin{itemize}
    \item Evaluation of a range of parameters for~\inplace~rotation and \teleturn~rotation
    \item Insights from our exploratory study (n=12) regarding user preference and performance with \inplace~rotation and \teleturn~rotation
\end{itemize}

\section{Related Work}

\subsection{Locomotion in VR}
To navigate through virtual environments (VEs), previous work proposed upper-body leaning~\cite{8809840} and several joystick- and keyboard-based locomotion techniques (e.g.,~\cite{bowmancontoller,ruddlekeyboard}).
However, physical movement through VEs has been largely shown to provide a better sense of orientation~\cite{chanceloco}, target direction~\cite{waller2004bodydirection}, and feeling of presence~\cite{usoh1999walking} even with a moderate level of visual detail~\cite{ruddleandlesselslod}.

Due to spatial limitations of the tracking space, variations of physical walking were proposed. While some approaches included treadmills (e.g.,~\cite{threadmill}), walking-in-place~\cite{slaterloco,templemanwip,10.1145/3013971.3014010} or jumping-in-place~\cite{jumpvr} to prevent users from reaching the physical boundaries of their tracking space, other techniques tried to ``trick'' the users' perception by using manipulations without being recognized~\cite{detectonthresholds}. Since visual calibration has been shown to occur when the presented self-speed was manipulated~\cite{mohlercalibration}, Williams et al. increased the users' reach by scaling the translational gain, i.e., the users' velocity~\cite{10.1145/1140491.1140495}. In a follow-up study, Williams et al. extended this technique by ``resetting'' the users' orientation when a physical boundary was reached, therefore forcing the users to physically reorient themselves towards the tracking space~\cite{10.1145/1272582.1272590}. Another variation of visual manipulation is redirected walking where users are nudged to walk on a circular path by virtually rotating their view (e.g.,\cite{razzaque2005redirected}). This technique, however, has been shown to be impractical in its naive implementation since the required walking radius would exceed the size of a typical tracking space~\cite{8613757}. To circumvent this limitation, resetting techniques and alternative approaches to substitute physical turns to keep the user within a limited physical space were proposed (e.g.~\cite{telewalk}).

Another well-explored locomotion technique is teleportation. According to the classification of locomotion techniques for VR by LaViola et al., teleportation is a selection-based travel technique~\cite{laviola20173d}. The users' viewpoint is instantly shifted to the destination that the user is pointing or looking at to prevent optical flow and the possibly accompanying motion sickness (e.g., \cite{bolte2011jumper,10.1145/2967934.2968105}). While some variants of teleportation limit the users to fixed teleportation destinations (i.e., fix-point teleportation), studies have shown that free teleportation leads to a lower discomfort~\cite{Frommel:2017:ECL:3102071.3102082}. Although teleportation can lead to disorientation~\cite{teleportdisorientation,teleportdisorientation2}, it was subjectively preferred over joystick input in a study by Langbehn et al. in 2018~\cite{langbehnteleportationvsjoystick}.

\subsection{Rotation in VR}
Physical rotation in VEs has been shown to be less time consuming~\cite{pauschsearchhead} and less error prone than controller-based rotation~\cite{avraamides2004headerror}.
A study by Ruddle and P{\'e}ruch, however, reported contradicting findings regarding the sense of direction when they compared physical and non-physical rotation in VEs~\cite{ruddle2004effects}.
Since physical rotation can be impractical due to a static body posture or tethering~\cite{Sargunam:2018:EJC:3191801.3191815}, there have been approaches to reduce or prevent physical rotation completely.
Similar to translational gains by Williams et al.~\cite{10.1145/1140491.1140495}, Kuhl et al. proposed to scale the users' physical rotation, therefore reducing the physical rotation needed for a full turn~\cite{kuhlscalingrotation}. This approach could be a benefit for setups with very limited tracking space (e.g., frontal tracking as used by the Oculus Rift CV1) or when tethering is an issue. Alternatively, Lin et al. propose to apply a smaller field-of-view (FOV) to reduce simulator sickness at the cost of user presence~\cite{FOV}. To combat this limitation and allow movement in a stationary position, Fernandes and Feiner suggest subtle dynamic changes in the users' FOV~\cite{dynamicfov}. However, a study by Sargunam et al. compared controller-based rotation techniques to prevent physical rotation and found that discrete rotation induced significantly less simulator sickness than reduced field-of-view and continuous rotation techniques~\cite{Sargunam:2018:EJC:3191801.3191815}.
Since a frame of reference (usually egocentric) and landmarks are very important for the users' orientation in VEs~\cite{mou2002intrinsic,foo2005landmarks,simulatedreferenceframe}, the question arises how large these discrete rotation intervals can become until users become disoriented. A study by Rahimi et al. applied discrete rotation between $45^\circ{}$ and $130^\circ{}$ without a significant increase in disorientation, however no evaluation of the effect of rotation magnitude was performed due to the randomized design of the experiment~\cite{rotationdiscreteanimatedteleport}. The underexplored impact of varying rotation parameters on user performance and presence led us to investigate a range of discrete rotation angles in a user study.

\section{User Experiment}
To evaluate how varying rotation parameters of the techniques~\inplace~and \teleturn~impact user performance, presence, and simulator sickness, a repeated measures 3 $\times$ 4 factorial design study was conducted. Rotation parameters included discrete rotation and the respective state-of-the-art for each technique.

\subsection{Method}
\label{sec:discrete_method}
Our within-subject experiment had two rotation techniques as independent variables: \inplace~and \teleturn. Both techniques were available in each condition and participants were free to choose one or alternate between them.\\
Rotation around the participants' position was defined by \inplace~and had three levels: physical rotation (\base), representing the current state-of-the-art, and two levels of discrete, button-based rotation at fixed angles (\ttwo~and \ffive). 
Rotation during the teleportation process was defined by \teleturn~and had four levels: no rotation (\none), which represents state-of-the-art teleportation from related work, \cont~rotation via circling the thumb on a touchpad, which is state-of-the-art in commercial VR games, and two levels of discrete, button-based rotation at fixed angles (\ttwo~and \ffive). The currently selected direction is indicated via a visual indicator (see Figure~\ref{fig:discrete_labyrinth} c).
Our experiment therefore consisted of $3 \times 4 = 12$ conditions that were fully counter-balanced.



\subsection{Implementation}
To explore the potential of controller-based discrete rotation, the study environment for the experiment was implemented for the Oculus Go HMD and its three DOF controller in Unity3D. Holding down the trigger button activated a teleportation parabola, releasing the button executed the teleportation to the selected destination point. For consistency, discrete rotation for both, \inplace~rotations (i.e., yaw axis rotations around the users' position) and \teleturn~(i.e., rotations during the teleportation process) was controlled by pressing either on the left or right side of the touchpad to initiate a rotation to the left or right, respectively. The angle of discrete rotation depended on the condition (either $22.5^\circ$ or $45^\circ{}$). Only during the \cont~\teleturn~condition, the relative position of the users' thumb on the touchpad defined the angle.
Similar to previous work, we opted for fractions of $90^\circ{}$ to allow fast $90^\circ{}$ rotations that were necessary to navigate an environment with orthogonal walls such as our labyrinth (see Figure~\ref{fig:discrete_labyrinth} a). The complexity of our level design was inspired by previous work. It has been shown that while a rotation oriented within-subject experiment requires an environment that enforces repeated rotations, a lack of complexity (e.g., simple rooms) leads to learning effects~\cite{rooms1,rooms2}.

\subsection{Participants}
We recruited 12 volunteers (6 male, 6 female) from our institution with a mean age of 22.75 (SD=2.38). All participants had normal or corrected-to-normal vision. Five reported previous VR experience and all but one reported to use physical rotation in VR instead of controller-based techniques.

\subsection{Measures}
After each condition, participants were asked to fill out questionnaires assessing their experience. Affective state was measured as valence, arousal, and dominance using the three 5-point pictorial scales of the the self-assessment manikin (SAM)~\cite{sam}. Simulator sickness was assessed with the Simulator Sickness Questionnaire (SSQ~\cite{ssq}, 16 items on a 4-point scale), while presence was measured with the igroup presence questionnaire (IPQ~\cite{ipq}, 14 items on a 7-point scale).
Objective measures logged during each condition were the number of teleportations, the number of rotations, the duration of teleportations (i.e., the time required to initiate and complete a teleportation including \teleturn~rotation), and the score achieved (number of items collected).
In a final questionnaire, participants were asked to rank all 12 conditions by preference and provide general comments.

\subsection{Procedure}
After an introduction, participants completed informed consent forms and provided information on their demographic background, including their VR experience. Afterwards, participants were explained the functionality of the HMD and the controller and had the opportunity to try the different rotation techniques. In each of the 12 conditions, participants were asked to navigate through a labyrinth via teleportation and collect as many items as possible within two minutes (see Figure~\ref{fig:discrete_labyrinth} b). In each condition, participants started at the center of the labyrinth. Items were spawned in a semi-random manner with equal difficulty. To estimate the users' disorientation, they were asked to return to the center of the labyrinth after each collected item, thereby gaining a score. Afterwards, they had to fill out the questionnaires and were asked to take a break to reduce carry-over effects of simulator sickness. All conditions were counter-balanced and only during conditions where \inplace~was set to \base~participants were allowed to rotate physically, otherwise rotation was performed via the controller. After the last condition, participants were asked to fill out the final questionnaire including a ranking of the conditions and general comments. Each participant received five currency (anonymized for review) of compensation at the end of the experiment.

\subsection{Results}
Due to the non-normal distribution of the collected data and multi-factorial design of the experiment, analysis was performed with the Aligned Rank Transform by Wobbrock et al. with Bonferonni correction for pairwise comparisons~\cite{art}. For readability's sake, descriptive statistics are presented in tables~\ref{tab:IPQSAM_descriptive} to ~\ref{tab:SSQ_performance_descriptive}. For clarity of presentation, only significant main effects and interactions will be reported below.

\begin{table*}[htb]
\centering
\small
    \begin{tabularx}{\textwidth}{l l *{16}{Y}}
    \toprule
    \multicolumn{2}{c}{\textsc{Condition}} & \multicolumn{2}{c}{\textsc{$IPQ_{G}$}} & \multicolumn{2}{c}{\textsc{$IPQ_{SP}$}} & \multicolumn{2}{c}{\textsc{$IPQ_{INV}$}} & \multicolumn{2}{c}{\textsc{$IPQ_{REAL}$}}  & \multicolumn{2}{c}{\textsc{$IPQ_{P}$}}\\
        \textit{InPlace} & \textit{TeleTurn} & \textit{M} & \textit{SD} & \textit{M} & \textit{SD} & \textit{M} & \textit{SD} & \textit{M} & \textit{SD} & \textit{M} & \textit{SD}\\
        \midrule
        \base & \none & 4.25 & 1.60 & 3.82 & 0.65 & 4.21 & 1.43 & 3.33 & 0.94 & 3.90 & 0.89\\
        \base & \cont & 4.17 & 1.34 & 3.43 & 0.77 & 3.85 & 1.23 & 2.96 & 1.10 & 3.60 & 0.87\\
        \base & \ttwo & 3.08 & 1.00 & 3.55 & 0.45 & 3.67 & 1.20 & 2.98 & 0.78 & 3.32 & 0.54\\
        \base & \ffive & 3.17 & 1.40 & 3.43 & 0.68 & 3.58 & 1.37 & 2.77 & 0.95 & 3.24 & 0.73\\
        \ttwo & \none & 4.17 & 1.34 & 3.82 & 0.76 & 4.17 & 1.17 & 3.17 & 1.02 & 3.83 & 0.96\\
        \ttwo & \cont & 3.83 & 1.03 & 3.63 & 0.62 & 4.15 & 1.05 & 2.79 & 0.97 & 3.60 & 0.59\\
        \ttwo & \ttwo & 3.58 & 1.31 & 3.67 & 0.51 & 3.92 & 1.38 & 3.08 & 1.11 & 3.56 & 0.70\\
        \ttwo & \ffive & 3.83 & 1.11 & 3.82 & 0.54 & 3.67 & 1.07 & 3.08 & 0.90 & 3.60 & 0.70\\
        \ffive & \none & 4.67 & 0.78 & 3.68 & 0.76 & 4.50 & 1.13 & 3.10 & 1.11 & 3.99 & 0.77\\
        \ffive & \cont & 3.83 & 1.27 & 3.63 & 0.43 & 4.06 & 1.01 & 3.00 & 0.87 & 3.63 & 0.40\\
        \ffive & \ttwo & 4.17 & 1.03 & 3.68 & 0.43 & 3.65 & 1.26 & 3.00 & 1.01 & 3.62 & 0.63\\
        \ffive & \ffive & 4.33 & .098 & 3.70 & 0.75 & 4.15 & 1.29 & 2.94 & 1.05 & 3.78 & 0.90\\
    \bottomrule
    \end{tabularx}
    \caption{Descriptive statistics by condition for the IPQ sub-scales general (\textsc{$IPQ_{G}$}), spatial (\textsc{$IPQ_{SP}$}), involvement (\textsc{$IPQ_{INV}$}), experienced realism (\textsc{$IPQ_{REAL}$}), and the IPQ presence score (\textsc{$IPQ_{P}$}).}
    \label{tab:IPQSAM_descriptive}
\end{table*}

\subsubsection{Presence}

\emph{IPQ Total Score}: There was a main effect for \inplace~($F_{2,22}=3.642$, \pequal .029) and \teleturn~($F_{3,33}=5.913$, \p .001). Post-hoc pairwise comparisons of \inplace~revealed that \ffive~(M=3.76, SE=0.17) was significantly higher than~\base~(\pequal .023, M=3.52, SE=0.18). Post-hoc pairwise comparisons of \teleturn~showed that \none~(M=3.91, SE=0.23) was significantly higher than~\ttwo~(\pequal .003, M=3.50, SE=0.16) and \ffive~(\pequal .002, M=3.54, SE=0.21).\\\\
\emph{IPQ General Item}: There was a main effect for \inplace~($F_{2,22}=3.236$, \pequal .043) and \teleturn~($F_{3,33}=5.911$, \p .001). Post-hoc pairwise comparisons of \inplace~revealed that \ffive~(M=4.25, SE=0.17) was significantly higher than~\base~(\pequal .043, M=3.67, SE=0.25). Post-hoc pairwise comparisons of \teleturn~revealed that \none~(M=4.36, SE=0.31) was significantly higher than~\ffive~(\pequal .008, M=3.79, SE=0.29) and~\ttwo(\pequal .001, M=3.61, SE=0.26).\\\\
\emph{IPQ Spatial Presence}: There were no main effects and no interaction.\\\\
\emph{IPQ Involvement}: There was a main effect for \teleturn~($F_{3,33}=4.146$, \pequal .008). Post-hoc pairwise comparisons of \teleturn~revealed that \none~(M=4.29, SE=0.32) was significantly higher than~\ttwo~(\pequal .013, M=3.74, SE=0.35) and \ffive~(\pequal .023, M=3.80, SE=0.32).\\\\
\emph{IPQ Expected Realism}: There was a main effect for \teleturn~($F_{3,33}=3.622$, \pequal .015).

\subsubsection{Affective State}
\emph{SAM Valence}: There was a main effect for \inplace~($F_{2,22}=5.892$, \pequal .003) and \teleturn~($F_{3,33}=9.390$, \p .001). Post-hoc pairwise comparisons of \inplace~revealed that \base~(M=3.40, SE=0.26) was significantly lower than~\ttwo\\(\pequal .008, M=3.96 SE=0.19) and \ffive~(\pequal .013, M=3.88, SE=0.17). Post-hoc pairwise comparisons of \teleturn~revealed that \cont~(M=3.31, SE=0.25) was significantly lower than~\ttwo~(\pequal .003, M=3.92, SE=0.17), \ffive~(\pequal .043, M=3.72, SE=0.24), and \none~(\p .001, M=4.03, Se=0.24).\\\\
\emph{SAM Arousal}: There was a main effect for \teleturn~($F_{3,33}=6.180$, \p .001). Post-hoc pairwise comparisons of \teleturn~revealed that \none~(M=3.36, SE=0.34) was significantly higher than \ttwo~(\p .001, M=2.83, SE=0.26) and \ffive~(\pequal .007, M=2.92, SE=0.29).\\\\
\emph{SAM Dominance}: There was a main effect for \inplace~($F_{2,22}=11.156$, \p .001) and \teleturn~($F_{3,33}=11.715$, \p .001). Post-hoc pairwise comparisons of \inplace~revealed that \base~(M=3.10, SE=0.23) was significantly lower than \ttwo (\p .001, M=3.79 SE=0.18) and \ffive~(\p .001, M=3.79, SE=0.17). Post-hoc pairwise comparisons of \teleturn~revealed that \none~(M=3.97, SE=0.26) was significantly higher than~\ttwo~(\pequal .009, M=3.64, SE=0.23), \ffive~(\pequal .005, M=3.50, SE=0.16), and \cont~(\p .001, M=3.14, Se=0.25). Furthermore, \ttwo~was significantly higher than \cont~(\pequal .041).

\begin{table*}[htb]
\centering
\small
    \begin{tabularx}{\textwidth}{l l *{16}{Y}}
    \toprule
    \multicolumn{2}{c}{\textsc{Condition}} &  \multicolumn{2}{c}{\textsc{$SAM_{V}$}} & \multicolumn{2}{c}{\textsc{$SAM_{A}$}} & \multicolumn{2}{c}{\textsc{$SAM_{D}$}}\\
        \textit{InPlace} & \textit{TeleTurn} & \textit{M} & \textit{SD} & \textit{M} & \textit{SD} & \textit{M} & \textit{SD}\\
        \midrule
        \base & \none & 4.08 & 1.08 & 3.58 & 1.38 & 4.17 & 1.11\\
        \base & \cont & 2.83 & 1.19 & 2.75 & 0.97 & 2.42 & 1.16\\
        \base & \ttwo & 3.33 & 0.78 & 2.92 & 0.90 & 3.00 & 1.21\\
        \base & \ffive & 3.33 & 1.15 & 2.83 & 1.03 & 2.83 & 1.03\\
        \ttwo & \none & 4.17 & 0.83 & 3.25 & 1.29 & 3.83 & 1.11\\
        \ttwo & \cont & 3.42 & 1.00 & 3.08 & 1.08 & 3.50 & 1.00\\
        \ttwo & \ttwo & 4.25 & 0.75 & 2.83 & 1.03 & 4.00 & 0.74\\
        \ttwo & \ffive & 4.00 & 0.85 & 2.83 & 1.03 & 3.83 & 0.72\\
        \ffive & \none & 3.83 & 0.94 & 3.25 & 1.22 & 3.92 & 1.24\\
        \ffive & \cont & 3.67 & 0.78 & 3.42 & 1.16 & 3.50 & 0.80\\
        \ffive & \ttwo & 4.17 & 0.72 & 2.75 & 1.22 & 3.92 & 0.90\\
        \ffive & \ffive & 3.83 & 0.94 & 3.08 & 1.24 & 3.83 & 0.58\\
    \bottomrule
    \end{tabularx}
    \caption{Descriptive statistics by condition for the SAM dimensions valence (\textsc{$SAM_{V}$}), arousal (\textsc{$SAM_{A}$}), and dominance (\textsc{$SAM_{D}$}).}
    \label{tab:SAM_descriptive}
\end{table*}

\begin{table*}[htb]
\centering
\small
    \begin{tabularx}{\textwidth}{l l *{16}{Y}}
    \toprule
    \multicolumn{2}{c}{\textsc{Condition}} & \multicolumn{2}{c}{\textsc{$SSQ_{N}$}} & \multicolumn{2}{c}{\textsc{$SSQ_{O}$}} & \multicolumn{2}{c}{\textsc{$SSQ_{D}$}} & \multicolumn{2}{c}{\textsc{$SSQ_{T}$}}\\
        \textit{InPlace} & \textit{TeleTurn} & \textit{M} & \textit{SD} & \textit{M} & \textit{SD} & \textit{M} & \textit{SD} & \textit{M} & \textit{SD}\\
        \midrule
        \base & \none & 11.93 & 12.29 & 15.80 & 13.11 & 18.56 & 24.71 & 17.45 & 14.99\\
        \base & \cont & 8.75 & 10.34 & 8.84 & 11.12 & 18.56 & 19.08 & 12.78 & 12.70\\
        \base & \ttwo & 7.16 & 9.21 & 13.27 & 13.38 & 11.60 & 20.42 & 12.47 & 14.73\\
        \base & \ffive & 8.75 & 16.50 & 15.16 & 24.83 & 17.40 & 47.90 & 15.58 & 30.25\\
        \ttwo & \none & 11.93 & 14.74 & 15.16 & 15.16 & 20.88 & 36.83 & 17.77 & 21.46\\
        \ttwo & \cont & 7.95 & 10.63 & 12.00 & 15.98 & 19.72 & 28.75 & 14.34 & 18.59\\
        \ttwo & \ttwo & 5.57 & 11.11 & 8.84 & 17.05 & 13.92 & 35.61 & 10.29 & 21.82\\
        \ttwo & \ffive & 3.98 & 11.11 & 7.58 & 12.93 & 13.92 & 31.41 & 9.04 & 18.90\\
        \ffive & \none & 11.93 & 13.57 & 13.27 & 12.57 & 15.08 & 15.08 & 15.27 & 13.76\\
        \ffive & \cont & 4.77 & 6.43 & 8.84 & 11.58 & 12.76 & 20.95 & 9.66 & 12.80\\
        \ffive & \ttwo & 5.57 & 13.77 & 11.37 & 21.56 & 15.08 & 35.34 & 11.84 & 25.31\\
        \ffive & \ffive & 5.57 & 11.83 & 11.37 & 23.86 & 15.08 & 40.02 & 11.84 & 26.68\\
    \bottomrule
    \end{tabularx}
    \caption{Descriptive statistics by condition for the SSQ sub-scales nausea (\textsc{$SSQ_{N}$}), oculomotor (\textsc{$SSQ_{O}$}), and disorientation (\textsc{$SSQ_{D}$}), and the total SSQ score (\textsc{$SSQ_{T}$}).}
    \label{tab:SSQ_descriptive}
\end{table*}

\subsubsection{Simulator Sickness}
\emph{SSQ Total Score}: There was a main effect for \teleturn\\($F_{3,33}=6.280$, \p .001). Post-hoc pairwise comparisons of \teleturn~revealed that \none~(M=16.83, SE=4.41) was significantly higher than \ttwo~(\pequal .013, M=11.53, SE=5.83) and \ffive~(\p .001, M=12.16, SE=7.15).\\\\
\emph{SSQ Nausea}: There was a main effect for \teleturn~($F_{3,33}=2.957$, \pequal .035). Post-hoc pairwise comparisons of \teleturn~revealed that \none~(M=11.93, SE=3.59) was significantly higher than \ffive~(\pequal .030, M=6.10, SE=3.45).\\\\
\emph{SSQ Oculomotor}: There was a main effect for \teleturn~($F_{3,33}=3.705$, \pequal .014). Post-hoc pairwise comparisons of \teleturn~revealed that \none~(M=14.74, SE=3.69) was significantly higher than \ffive~(\p .014, M=11.37, SE=5.75) and \cont~(\pequal .048, M=9.90, SE=3.10).\\\\
\emph{SSQ Disorientation}: There was a main effect for \teleturn~($F_{3,33}=2.983$, \pequal .034). Post-hoc pairwise comparisons of \teleturn~revealed that \cont~(M=17.01, SE=5.81) was significantly higher than \ffive~(\pequal .045, M=15.74, SE=11.40).

\begin{table*}[htb]
\centering
\small
    \begin{tabularx}{\textwidth}{l l *{16}{Y}}
    \toprule
    \multicolumn{2}{c}{\textsc{Condition}} & \multicolumn{2}{c}{\textsc{Teleport}} & \multicolumn{2}{c}{\textsc{Duration}} & \multicolumn{2}{c}{\textsc{$Rotations_{IP}$}} & \multicolumn{2}{c}{\textsc{$Rotations_{TT}$}}\\
        \textit{InPlace} & \textit{TeleTurn} & \textit{M} & \textit{SD} & \textit{M} & \textit{SD} & \textit{M} & \textit{SD} & \textit{M} & \textit{SD}\\
        \midrule
        \base & \none & 97.17 & 40.50 & 487 & 348 & -- & -- & -- & --\\
        \base & \cont & 57.67 & 20.46 & 1427 & 518 & -- & -- & 32.33 & 13.43\\
        \base & \ttwo & 60.00 & 27.48 & 1754 & 1042 & -- & -- & 115.33 & 23.64\\
        \base & \ffive & 62.25 & 25.59 & 1507 & 792 & -- & -- & 77.75 & 27.68\\
        \ttwo & \none & 78.33 & 30.93 & 538 & 398 & 165.42 & 51.58 & -- & --\\
        \ttwo & \cont & 69.17 & 25.32 & 741 & 404 & 109.58 & 56.65 & 14.83 & 9.93\\
        \ttwo & \ttwo & 71.83 & 34.73 & 1884 & 1686 & 131.58 & 59.22 & 39.25 & 53.88\\
        \ttwo & \ffive & 65.25 & 25.69 & 869 & 336 & 88.58 & 40.94 & 31.08 & 38.83\\
        \ffive & \none & 89.58 & 41.91 & 452 & 336 & 127.08 & 39.79 & -- & --\\
        \ffive & \cont & 66.33 & 26.43 & 997 & 787 & 84.33 & 44.34 & 14.67 & 13.23\\
        \ffive & \ttwo & 71.83 & 37.05 & 749 & 434 & 134.00 & 65.02 & 28.50 & 34.05\\
        \ffive & \ffive & 72.00 & 23.28 & 679 & 480 & 97.42 & 40.23 & 23.83 & 34.74\\
    \bottomrule
    \end{tabularx}
    \caption{Descriptive statistics by condition for the number of teleportations (\textsc{Teleport}), duration of teleportation (in ms), number of discrete rotations in place ($Rotations_{IP}$), and number of discrete rotations during teleportations ($Rotations_{TT}$).}
    \label{tab:SSQ_performance_descriptive}
\end{table*}
\subsubsection{Number of Teleportations}
There was a main effect for \teleturn~($F_{3,33}=9.774$, \p .001). Post-hoc pairwise comparisons of \teleturn~revealed that \none~(M=88.36, SE=9.69) was significantly higher than \ttwo~(\p .001, M=67.89, SE=8.81), \ffive~(\p .001, M=66.50, SE=6.17), and \cont~(\p .001, M=64.39, SE=6.13).

\subsubsection{Duration of Teleportation}
A two-way repeated measures ANOVA with\\ Greenhouse-Geisser correction revealed a significant main effect for \inplace~on the participants' teleportation duration (\F{1.30}{14.25} 9.56, \pequal .005, \etasq .47, $\epsilon$=.648). Pairwise t-tests with Bonferonni correction revealed that the duration of teleportation was significantly higher in the \inplace~\base~condition than in the \inplace~\ffive~condition (\p .001). In addition, there was a significant main effect for \teleturn~(\F{1.71}{18.83} 11.57, \pequal .001, \etasq .51, $\epsilon$=.571). Pairwise t-tests with Bonferonni correction revealed that the duration of teleportations was significantly lower in the \teleturn~\none~condition than in the \teleturn~\cont~(\pequal .001), \teleturn~\ttwo~(\pequal .005), and \teleturn~\ffive~condition (\pequal .001). Furthermore, there was a significant interaction between \inplace~and \teleturn~(\F{2.61}{28.75} 4.80, \pequal .010, \etasq .30, $\epsilon$=.436). While teleportation duration was mostly unaffected by the \inplace~level during \teleturn~\none~conditions, teleportation duration in the \teleturn~\ttwo~and \teleturn~\ffive~conditions was lowest with \inplace~levels at \ffive.

\subsubsection{Number of Discrete Rotations}
Rotations were divided into rotations that happened in place, i.e. around the participants' position, ($Rotations_{IP}$) and rotations during teleportation, i.e. rotating the indicator arrow during teleportation, ($Rotations_{TT}$, see Figure~\ref{fig:discrete_labyrinth} c).
A Wilcoxon Signed Rank test revealed that the $Rotations_{IP}$ ranks for the \inplace~\ttwo~condition (mean rank = 27.04) were significantly higher than in the \inplace~\ffive~condition (mean rank = 23.46, \Z -2.15, \pequal .032). Median (IQR) $Rotations_{IP}$ levels for the \inplace~\ttwo~and \inplace~\ffive~condition were 127.00 (81.50 to 163.00) and 117.50 (78.50 to 139.00), respectively.

A Friedman's ANOVA revealed significant differences in the $Rotations_{TT}$ values for the three \teleturn~conditions \teleturn~\cont, \teleturn~\ttwo, and \teleturn~\ffive~(\Friedman(2)=9.31, \pequal .009). Median (IQR) $Rotations_{IP}$ levels for the \teleturn~\cont, \teleturn~\ttwo, and \teleturn~\ffive~conditions were 16.00 (7.25 to 32.00), 58.50 (1.50 to 112.75), and 43.50 (1.00 to 82.50), respectively. Wilcoxon Signed Rank tests were performed with a Bonferonni adjusted p-value of .017. $Rotations_{TT}$ ranks were significantly higher in the \teleturn~\ttwo~condition (mean rank = 22.78) than in the \teleturn~\cont~condition (mean rank = 8.77, \Z -3.12, \p .001) and $Rotations_{TT}$ ranks were significantly higher in the \teleturn~\ttwo~condition (mean rank = 18.59) than in the \teleturn~\ffive~condition (mean rank = 11.17, \Z -3.06, \pequal .002). Furthermore, $Rotations_{TT}$ ranks were significantly higher in the \teleturn~\ffive~condition (mean rank = 24.07) than in the \teleturn~\cont~condition (mean rank = 9.75, \Z -3.09, \pequal .002).

\begin{figure}[htb]
    \centering
    \begin{subfigure}[t]{.7\textwidth}
    \centering
    \subcaption{}
        \includegraphics[width=\textwidth]{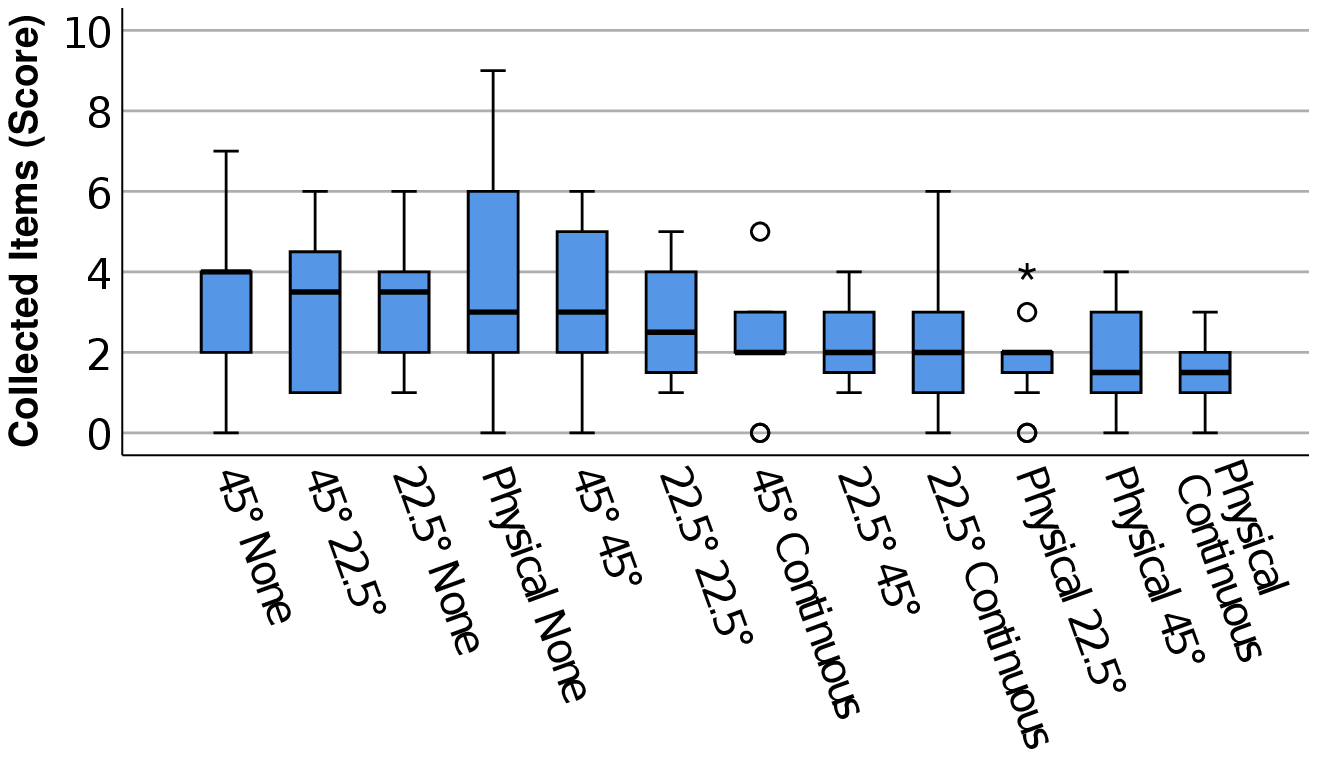}
       
    \end{subfigure}
    \begin{subfigure}[t]{.7\textwidth}
    \centering
    \subcaption{}
        \includegraphics[width=\textwidth]{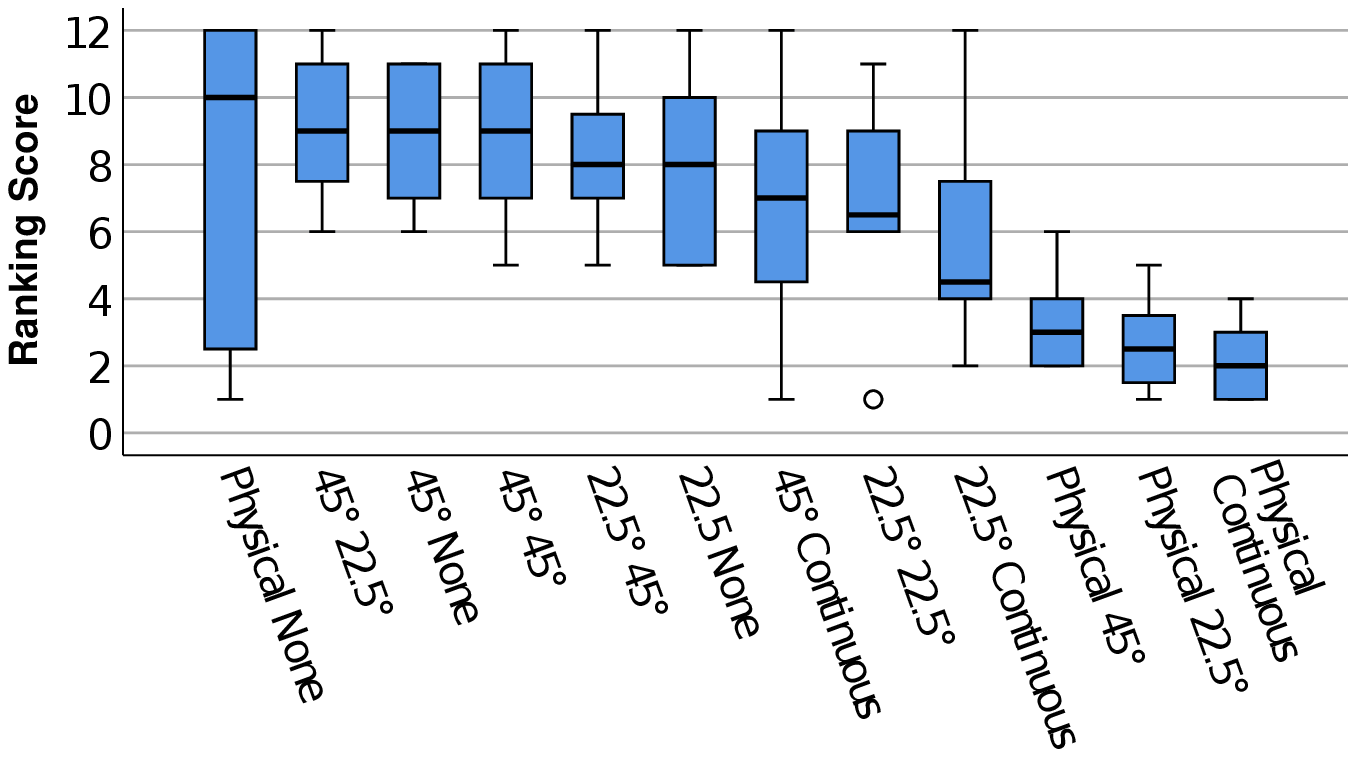}
    \end{subfigure}
    \caption{Participants' score (a) and ranking of the conditions (b) in descending order.}
    \label{fig:score_and_ranking}
\end{figure}

\subsubsection{Participant Score}
There was a main effect for \inplace~($F_{2,22}=4.188$, \pequal .017) and \teleturn~($F_{3,33}=5.306$, \pequal .002). Post-hoc pairwise comparisons of \inplace~revealed that \ffive~(M=3.00, SE=0.30) was significantly higher than~\base~(\pequal .013, M=2.23, SE=0.26). Post-hoc pairwise comparisons of \teleturn~revealed that \none~(M=3.47, SE=0.45) was significantly higher than \cont~(\p .001, M=1.92, SE=0.24). On average, the \base~\none~condition achieved the highest and the \base~\cont~condition the lowest score (see Figure~\ref{fig:score_and_ranking}).

\subsubsection{Ranking of Conditions}
There was a main effect for \inplace\\($F_{2,22}=42.439$, \p .001) and \teleturn~($F_{3,33}=9.452$, \p .001). Post-hoc pairwise comparisons of \inplace~revealed that \ffive~(M=8.38, SE=0.35) was significantly higher than \ttwo~(\pequal .047, M=7.23, SE=0.34) and \base~(\p .001, M=3.90, SE=0.30). Furthermore, \ttwo was significantly higher than \base~(\p .001). Post-hoc pairwise comparisons of \teleturn~revealed that \none~(M=8.17,\\SE=0.69) was significantly higher than \ttwo~(\pequal .043, M=6.25, SE=0.36) and \cont~(\p .001, M=4.81, SE=0.60). Furthermore, \ffive~was significantly higher than \cont~(\pequal .002). On average, the \ffive~\ttwo~condition achieved the highest rank, while the \base~\cont~condition achieved the lowest rank (see Figure~\ref{fig:score_and_ranking}).

\section{Discussion}
\subsection{\inplace}
Our results indicate that independent of the rotation interval, participants felt more pleased and in control with discrete \inplace~rotations (i.e., \inplace~conditions \ttwo~ and \ffive) than with physical rotation. Even the largest rotation interval did not affect simulator sickness and spatial disorientation levels. In contrast, participants were able to navigate the labyrinth more efficiently, which is reflected in the higher number of collected items. Both discrete \inplace~conditions were preferred over physical rotation, with the highest rotation interval (\ffive) being the favorite. Although rotations at \ffive~were considered ``easier'' (P7), they were not fast enough for some participants, calling for a ``higher rotation interval'' (P4) or the possibility to ``hold down the rotation button'' (P12). Higher rotation intervals should therefore be considered in future studies of \inplace~rotation.\\\\
\textbf{Implications:} Some participants considered repeated button presses for both rotation types as too slow, while others considered the rotation intervals as too large and imprecise.
Depending on the user preference and environment, it should be possible to adjust the rotation interval similar to adjusting the scrolling speed of a computer mouse.

\subsection{\teleturn}
Since \teleturn~is a novel interaction technique included in a limited set of commercial VR games, we included the \none~condition, which represents teleportation from related work (e.g., \cite{teleportation}), in addition to the \cont~state-of-the-art in our user experiment. While participants preferred discrete \teleturn~rotations over the \cont~condition and felt more pleased and in control, the \teleturn~\none~condition was superior in terms of interaction time (teleportation duration) and performance (participant score). This could be explained by the lack of experience regarding \teleturn~rotation. P9 considered \teleturn~rotation as ``tedious'' and preferred a ``sequence'' of \inplace~rotation and straight teleportation. This is consistent with the number of rotations observed. In conditions with discrete \inplace~rotation, the number of \teleturn~rotations was lower since participants were free to choose between either of the rotation modes.\\\\
\textbf{Implications:} Participants mostly preferred rotation around their position (\inplace) over rotations during teleportation (\teleturn), since the technique was ``intuitive to use [...] but difficult to get used to'' (P5). While a preview of the new orientation direction in form of a visual indicator was helpful, some users might benefit from a preview of the perspective to be assumed in order to see a benefit in \teleturn~over \inplace~rotation. Similar to Liu et al., a portal or a video screen floating next to the user could serve as a medium to achieve this effect~\cite{portal}. Furthermore, the \teleturn~condition was new to most participants, suggesting that interaction time could be reduced due to learning effects, making the technique more efficient. Since the current experimental setup set a time restriction, participants could have been in favor of the fastest technique. Considering other applications, such as shooters, where teleporting blindly to a new position might be dangerous, the time needed to select a new orientation via \teleturn~might be put into a new perspective.

\subsection{Summary}
Overall, teleportation with physical rotation (\base~\none~condition), which is the current state-of-the-art of most VR applications that feature teleportation, achieved the highest ranking score. A combination of \ffive~\inplace~rotation and no \teleturn~rotation (\none) achieved the highest score and was the third most preferred condition, offering a viable alternative for scenarios where physical rotation is impractical or impossible (e.g., sitting or lying). The novel \teleturn~technique with \cont~rotation in combination with physical \inplace~rotation, as it is implemented in some commercial VR games, achieved the lowest score and was the least preferred condition of our participants. Since a combination of \ffive~\inplace~and \ttwo~\teleturn~rotation achieved the second highest score and was the second most preferred condition, we consider that \teleturn~could successfully be improved over its \cont~state-of-the-art and was well accepted by our participants.

\section{Limitations}
The experiment was performed with a three DoF HMD. Nevertheless we argue that the results can also be transferred to six DOF tracking, since the focus of the study was on rotation and not translation.
The small sample size in our experiment could have been too low to find small effects between conditions and we explored only a limited range of rotation angles that might have been in favor of our game level design. Values beyond the investigated range could yield other results in terms of simulator sickness and player performance. Furthermore, spatial orientation was measured via the number of collected items and the respective SSQ subscale. Alternatively, orientation could have been measured via spatial updating which could result in a different outcome~\cite{spatialupdating}.

\section{Conclusion}
In this work we investigated the nature and potential of discrete rotation in VR to avoid physical rotation and maintain user presence, performance, and prevent simulator sickness. Two rotation techniques, rotation in-place (\inplace) and rotation during the teleportation process (\teleturn), were evaluated with fixed rotation intervals against their respective state-of-the-art. Our results indicate that discrete \inplace~rotation was preferred over physical rotation and significantly improved user presence and performance. 
The discrete \teleturn~variation was preferred over continuous \teleturn, which is the current state-of-the-art, and led to significantly lower disorientation. \inplace~rotation was largely preferred over \teleturn. As a result, participants adapted a locomotion style that combined discrete \inplace~rotations with straight teleportation without \teleturn~suggesting a complementary usage. From our findings we conclude that teleportation without physical rotation is feasible while maintaining user experience and performance.

\bibliographystyle{ACM-Reference-Format}
\bibliography{main}

\end{document}